\documentclass{iau}

\usepackage{amsmath}
\usepackage{graphicx}
\usepackage{multirow}

\begin{document}

\lefttitle{Berriman}
\righttitle{Astronomy in Focus}

%\jnlPage{1}{7}
%\jnlDoiYr{2021}
%\doival{10.1017/xxxxx}

%\pubyear{2025}
\setcounter{page}{1}
%\jname{Astronomy in Focus, Volume 1} 

\aopheadtitle{Astronomy in Focus}
\editors{ Y. Azzam, Z. BenKhaldoun, D. Buckly, and M. Povic eds.}

\title{The Benefits of the Virtual Observatory to Underserved Communities}

\author{G. Bruce Berriman}
\affiliation{Caltech/IPAC-NExScI, \\ 1200 East California Boulevard, Pasadena, CA~~91125, USA}

\begin{abstract}
The Virtual Observatory (VO) is a global ecosystem of inter-operating services that connect worldwide data archives. The VO is implemented in all major astronomy archives through common interfaces developed by the 22 members of the International Virtual Observatory Alliance (IVOA). It was founded in 2002, and the newest members, the SKA Observatory and the Kazakhstan Virtual Observatory, joined in 2022. The VO offers access to data on FAIR principles and from its inception has supported Open Science. The VO acts as a democratizing influence in astronomy: it provides equal access to worldwide public data sets to under-served communities as well as to large data centers, and it enables international participation in scientific research and education. Thus, astronomers from many different communities are positioned to participate in the big science questions emerging in astronomy in the 2020s, such as interpreting transient sources that will be measured in forthcoming missions such as Rubin. In addition, the IVOA has signed an MoU with the IAU Office of Astronomy for Development (OAD). Under this MoU, IVOA members participated in "Astronomy from Archival Data," which involved educational activities for undergraduate and post-graduate students organized by Dr. Priya Hasan. The IVOA plans to participate in future such educational events. The presentation describes how new communities may participate in Virtual Observatory science and educational activities, including practices for developing VO-compliant data centers and archives and education and training for developers and end users.

\end{abstract}


\maketitle

\section{Introduction: What Is the Virtual Observatory?}

The Virtual Observatory (VO) is an ecosystem of inter-operating tools and services deployed by distributed data centers and archives. It can be thought of as a multi-wavelength digital sky. Architecturally, the VO is implemented as a layer of services built into archives and data centers, that enable  discovery, access, exploration and analysis of worldwide data in a uniform fashion. VO services are being actively embedded in the architectures of new missions and observatories, such as Rubin, Roman, and Euclid, and are central to their future operations \citep{O'TooleSPIE}.

Since its inception, the VO has been built on the principle of open access and inclusivity (except for data subject to periods of exclusive use). The VO acts as a democratizing force in astronomy, in that data may be accessed by anyone with an internet connection. Astronomers and educators in under-served communities and developing countries are now taking part in VO development work and taking advantage of the VO in astronomy research.

\section{The Role of the IVOA in the Virtual Observatory}

The VO is built on metadata standards and data models developed by the International Virtual Observatory Alliance (IVOA) (\verb"https://ivoa.net") expressly to offer uniform access to data. Data centers and archives build and operate the software services that implement these standards.

The IVOA itself was founded in 2002 \citep{2004IVO} by three organizations, and this number has grown to 24 in 2024. Membership is open to national VO projects and to intergovernmental organizations (IGOs), who seek funding from their respective funding agencies. Member organizations have a seat on the Executive Committee, which has has overall responsibility for management of the IVOA.  

The two newest members of the IVOA reflect the diversity of organizations within the alliance. The Square Kilometer Array Observatory (SKAO; \verb"https://www.skao.int/en") is an IGO that represents a global collaboration of states who will operate the Square Kilometer Array (SKA) telescope. The Kazakhstan VO (KVO; \verb"https://vo.fai.kz/") will develop the VO in the Republic of Kazakhstan. The KVO is a particularly good example of contributions made by developing countries. Its archive will be built on VO standards and will incorporate data from robotic telescopes located at the Assy-Turgen Plateau, with data processed and managed by a powerful computational cluster. The archive will include a digitized library archive of about 20,000 photographic plates \citep{2022NewA...9701881S}. 

\section{The IVOA Development Model}

The IVOA's work focuses on semi-annual hybrid "interoperability workshops," one in the Northern Spring of each year and one in the Northern Fall, usually consecutive with the annual ADASS meeting. Attendance is unrestricted and not limited to persons associated with member organizations. The IVOA seeks full sponsorship for the meetings to avoid registration fees and encourage attendance. Four meetings held virtually during the COVID pandemic attracted nearly 200 registrants each, approximately 35\% more than for in-person meetings. For this reason, post-pandemic meetings are now hybrid as a matter of course. The last three meetings have been in Bologna, Italy; Tucson, AZ, USA; and Sydney, Australia. The next meeting is in Valleta, Malta in November 2024. Meeting proceedings can be found at \verb"https://wiki.ivoa.net/twiki/bin/view/IVOA/IvoaEvents". 

The IVOA follows a formal process for approving standards. They are developed by dedicated working groups, whose roles are to achieve consensus on the content of protocols and to document them for review by the broader community. The standards themselves respond to science drivers recommended by the Committee for Science Priorities (CSP). After a six-week Request for Comment period, the standards are revised and approved by the working groups' coordinating body, Technical Working Group (TCG). The CSP, TCG, and working groups are all overseen by the Executive Committee, which gives final approval of standards submitted to it by the TCG (see \verb"https://ivoa.net/documents/index.html#proc").

\section{The VO and FAIR Principles}

Making data open does not by itself ensure they are discoverable and sustainable. The IVOA therefore actively engages with the international scientific community to ensure that astronomy follows emerging practices for managing metadata and data. An outstanding instance of this is to ensure, as far as is as possible, that data comply with the fifteen Findable, Accessible, Interoperable, and Reusable (FAIR) principles enumerated in \cite{2016Scidata}. The principles are intended to promote the mechanics of machine access to metadata. Compliance with FAIR principles is an international effort across research infrastructures in many disciplines, often at the direction of funding agencies. The IVOA has already made considerable progress in this regard \citep{2024ASPC..535..265O}; VO standards are in fact in line with 10 of the 15 principles, though no formal verification process for compliance has yet been applied. Some are out of scope for the IVOA, while others are under active development. At the request of the IAU, the IVOA has begun engaging with the Cross-Domain Interoperability Framework (CDIF) project \citep{CDIF2023} under the auspices of CODATA (\verb"https://codata.org/") to share expertise and practices. 

\section{VO Tools and Services}

There are many open tools available that aggregate and display multiple catalog, imaging, and spectroscopic data sets that are served through VO protocols. There is a complete list at \verb"https://ivoa.net/astronomers/applications.html". Four of the most widely used tools are shown in Table \ref{tab:Table 1}. They are all actively updated to incorporate changes in the astronomy data landscape.

\begin{table} [h!]
    \centering
    \begin{tabular}{ll}
      {\bf Tool}   & {\bf URL}                                          \\
        Aladin     & https://aladin.cds.unistra.fr/                     \\
        TOPCAT     & https://www.star.bris.ac.uk/~mbt/topcat/          \\
        ESASky     & https://sky.esa.int/esasky/                         \\
        Firefly    & https://github.com/Caltech-IPAC/firefly            \\
                 &                                                       \\
    \end{tabular}
    \caption{Four widely used VO tools for discovering and aggregating data made accessible through the VO.}
    \label{tab:Table 1}
\end{table}
\section{Data Availability}

Astronomers have taken full advantage of freely available data sets. An outstanding example is the data released by the Gaia mission, which are only accessible through the VO. The community has taken full advantage of these data to publish a total of over 9,000 papers since 2015 (as revealed by searches of the Astrophysics Data System), many from astronomers from developing countries; e.g., \cite{2024MNRAS.531.3715A}, \cite{2024NewA..10902196B}, \cite{2022JAsGe..11..142B}, \cite{2024NewA..11002231H}, \cite{2024A&A...686A.225I}, \cite{2024arXiv240811383T}, and \cite{2022AdSpR..69..467T}.

\section{Engagement of the IVOA with the IAU}
Engagement with the IAU is a central goal  of the IVOA through the 2020s. In 2021, the IVOA signed a Memorandum of Understanding (MoU) with the IAU Office of Astronomy and Development (OAD) to use VO tools and services to foster inclusiveness and education, and to foster global development. These are two of the goals in the IAU Strategic plan: \verb"https://www.iau.org/administration/about/strategic_plan/". The MoU came about because of the success of the "Astronomy from Archival Data" project that engaged 915 persons from 23 countries \citep{2021IAUS..367..409H} over six months. IVOA members took part in teaching and mentoring, which included demonstrations of use cases that involved the tools in Table \ref{tab:Table 1}. A total of 80 videos from the course have been posted on YouTube at \verb"https://tinyurl.com/bdhwk49h".  

Another development in 2021 was the establishment within Division B of a VO Functional Working Group. This VO Working Group organized a session of five talks at the 2024 IAU General Assembly on "Community Engagement, Open Science and the Virtual Observatory." The presentations included one on the creation of a manual of science use cases created in a course for disadvantaged students presented at Copperhead University, Zambia. \citep{2024IAUOpenScience}. Finally, a side event at the General Assembly was a workshop on "VO Training for African Students." \verb"https://www.overleaf.com/project/664da7dda8952777f0931539" describes the content of the workshop. Again, the tools listed in Table \ref{tab:Table 1} were widely used in the workshop.

\section{Educational Resources} 
There are many many freely available educational resources, usually from summer schools and workshops held by the IVOA. These are posted on the IVOA education website (\verb"https://ivoa.net/documents/Notes/EDU/index.html"). A semester-long course on the VO, developed by the University of Heidelberg, is available at \verb"https://codeberg.org/msdemlei/vo-course.git." 
There are fewer resources for training developers; the principal resource is "Publishing in the VO," which is available at \verb"https://tinyurl.com/3mu8zu6m". 

\section{Participation in the IVOA}
The IVOA faces many challenges going forward: support for very large data sets and science platforms; support for new missions; follow-up to transient events reported by Rubin, including networks of ground-based telescopes or follow-up; and making legacy data sets available. All persons are free to participate in the IVOA's activities and register for IVOA meetings. Participation can include developing, evaluating, and documenting standards; developing new use cases; and supporting web page updates. A twice-nearly newsletter reports on IVOA standards, tools, and science results. An active X account and Slack promote discussion of matters of immediate interest.

\end{document}